\documentclass{aa}

\usepackage{graphicx}
\usepackage{txfonts}
\usepackage[print-unity-mantissa=false]{siunitx} 
\usepackage[dvipsnames]{xcolor}
\usepackage{float}
\usepackage{ulem}
\usepackage{natbib,twoopt}
\usepackage[breaklinks=true]{hyperref} 
\bibpunct{(}{)}{;}{a}{}{,}             
\makeatletter
  \newcommandtwoopt{\citeads}[3][][]{\href{http://adsabs.harvard.edu/abs/#3}%
    {\def\hyper@linkstart##1##2{}%
     \let\hyper@linkend\@empty\citealp[#1][#2]{#3}}}
  \newcommandtwoopt{\citepads}[3][][]{\href{http://adsabs.harvard.edu/abs/#3}%
    {\def\hyper@linkstart##1##2{}%
     \let\hyper@linkend\@empty\citep[#1][#2]{#3}}}
  \newcommandtwoopt{\citetads}[3][][]{\href{http://adsabs.harvard.edu/abs/#3}%
    {\def\hyper@linkstart##1##2{}%
     \let\hyper@linkend\@empty\citet[#1][#2]{#3}}}
  \newcommandtwoopt{\citeyearads}[3][][]%
    {\href{http://adsabs.harvard.edu/abs/#3}
    {\def\hyper@linkstart##1##2{}%
     \let\hyper@linkend\@empty\citeyear[#1][#2]{#3}}}
\makeatother
\usepackage{cleveref}
\usepackage{siunitx}
\usepackage{upgreek}
\DeclareSIUnit{\pc}{\text{pc}}
\DeclareSIUnit{\yr}{\text{yr}}
\DeclareSIUnit{\kpc}{\kilo\pc}
\DeclareSIUnit{\solmass}{\mathrm{M_{\odot}}}
\DeclareSIUnit{\gyr}{\text{Ga}}

\crefname{equation}{Eq.}{Eqs.}
\Crefname{equation}{Equation}{Equations}
\crefname{figure}{Fig.}{Figs.}
\Crefname{figure}{Figure}{Figures}
\crefname{section}{Sect.}{Sects.}
\Crefname{figure}{Section}{Sections}
\newcommand{\md}{\mathrm{d} }
\newcommand{\deriv}[2]{\dfrac{\mathrm{d} #1}{\mathrm{d} #2}}

\newcommand{\sod}{\sigma_{\rm 1D}}
\newcommand{\sbm}{\sigma/m_{\chi}}
\newcommand{\nbody}{$N$-body }
\newcommand{\rhosp}{\rho_{\rm sp}}

\newcommand{\rsp}{r_{\rm sp}}
\newcommand{\rmax}{r_{\rm max}}
\newcommand{\rmin}{r_{\rm min}}

\newcommand{\risco}{r_{\rm ISCO}}
\newcommand{\mbh}{M_{\rm BH}}

\newcommand{\lcdm}{$\rm \Lambda CDM$ }
\newcommand{\medd}{\dot{M}_{\rm Edd}}

\newcommand{\dd}{\mathrm{d}}

\begin{document} 
   \title{Accretion of self-interacting dark matter onto supermassive black holes}
    \author{Sabarish V. M.\inst{1}
          \and
          Marcus Brüggen\inst{1}
          \and
          Kai-Schmidt Hoberg\inst{2}
          \and
          Moritz S.\ Fischer\inst{3,4}
          }
   \institute{
    Hamburger Sternwarte, Universität Hamburg, Gojenbergsweg 112, D-21029 Hamburg, Germany\\
              \email{sabarish.venkataramani@uni-hamburg.de}
    \and
    Deutsches  Elektronen-Synchrotron  DESY,  Notkestr.~85,  22607  Hamburg,  Germany
    \and 
    Universit\"ats-Sternwarte, Fakult\"at für Physik, Ludwig-Maximilians-Universit\"at M\"unchen, Scheinerstr.~1, D-81679 M\"unchen, Germany
    \and
    Excellence Cluster ORIGINS, Boltzmannstrasse 2, D-85748 Garching, Germany
    }


  \abstract
   {Dark matter (DM) spikes around supermassive black holes (SMBHs) may lead to interesting physical effects such as enhanced DM annihilation signals or dynamical friction within binary systems, shortening the merger time and possibly addressing the `final parsec problem'. They can also be promising places to study the collisionality of DM because their velocity dispersion is higher than in DM halos allowing us to probe a different velocity regime.
   }
   {We aim to understand the evolution of isolated DM spikes for collisional DM and compute the BH accretion rate as a function of the self-interaction cross-section.
   }
   {We have performed the first $N$-body simulations of self-interacting dark matter (SIDM) spikes around supermassive black holes (SMBH) and studied the evolution of the spike with an isolated BH starting from profiles similar to the ones that have been shown to be stable in analytical calculations.}
   {We find that the analytical profiles for SIDM spikes remain stable over the time-scales of hundreds of years that we have covered with our simulations. In the long-mean-free-path (LMFP) regime, the accretion rate onto the BHs grows linearly with the cross-section and flattens when we move towards the short-mean-free-path (SMFP) regime. In both regimes, our simulations match analytic expectations, which are based on the heat conduction description of SIDM. A simple model for the accretion rate allows us to calibrate the heat conduction in the gravothermal fluid prescription of SIDM. Using this prescription, we determine the maximum allowed accretion rate which occurs when $\risco \rho(\risco) \sbm \sim 1$, where $\sbm$ is the self-interaction cross-section and $\risco$ the radius of the innermost stable orbit. 
    }
   {Our calibrated DM accretion rates could be used for statistical analysis of SMBH growth and incorporated into subgrid models to study BH growth in cosmological simulations.}

   \keywords{dark matter -- black hole physics -- methods: numerical
               }

   \maketitle
\section{Introduction}
Self-interacting dark matter (SIDM) models are a class of models where dark matter (DM) can scatter off each other non-gravitationally \citep{spergelObservationalEvidenceSelfInteracting2000}. 
SIDM has been invoked to solve potential small-scale problems of $\Lambda$CDM cosmology \citep{bullockSmallScaleChallengesLCDM2017}. 
The most robust constraints on the self-interacting cross-section come from observations of galaxy clusters. 
In particular, by measuring the central density profiles of galaxy clusters  \citep[e.g.][]{sagunskiVelocitydependentSelfinteractingDark2021, andradeStringentUpperLimit2021}, and merging clusters such as the bullet cluster \citep{randallConstraintsSelfInteractionCross2008,markevitchDirectConstraintsDark2004}. 

The effects of SIDM on astrophysical systems have been studied widely using \nbody simulations of systems ranging from isolated dwarf galaxies \citep[e.g.][]{yangGravothermalEvolutionDark2022} over merging galaxy clusters \citep[e.g.][]{aridoSimulatingRealisticSelfinteracting2025, sabarishSimulationsGalaxyCluster2024,kimWakeDarkGiants2017,fischerUnequalmassMergersDark2022} to cosmological simulations \citep{correaTangoSIDMTantalizingModels2022,fischerCosmologicalSimulationsRare2022, ragagninDianogaSIDMGalaxy2024,despaliIntroducingAIDATNGProject2025}. 
In addition, semi-analytic models exist to understand the evolution of SIDM halos \citep[e.g.][]{balbergSelfInteractingDarkMatter2002, jiangSemianalyticStudySelfinteracting2023} and are powerful tools to constrain SIDM models \citep[e.g.][]{yangConstrainingSIDMCross2023}.

On even smaller scales, it is expected that we find dense DM distributions around black holes (BH) \citep{gondoloDarkMatterAnnihilation1999,alachkarDarkMatterConstraints2023,chanFirstRobustEvidence2024}. 
The density distribution has a power-law profile $\rho(r) \propto r^{-\gamma}$, and is dubbed a DM spike. In the case of cold dark matter (CDM) it was shown theoretically that an adiabatic contraction of the DM halo that hosts the BH could lead to the formation of a DM spike \citep{sadeghianDarkmatterDistributionsMassive2013, gondoloDarkMatterAnnihilation1999}. 
For CDM, the spike index can be estimated from the initial CDM profile in which 
the BH is found. 
The resulting index is $\gamma = (9-2\beta)/(4-\beta)$, where $\beta$ is the index of the initial density profile in the inner regions \citep{gondoloDarkMatterAnnihilation1999}, that is $\rho \propto r^{-\beta}$. 
If the initial DM density follows a Navarro-Frenk-White (NFW) \citep{navarroStructureColdDark1996} profile, a spike index of $\gamma=7/3$ is obtained using $\beta=1$. 
Therefore, such DM spikes are expected to be found around supermassive black holes (SMBH) since they are typically found inside a galactic halo. More recently, it was suggested that DM forms shallower mounds instead of denser spikes due to different assumptions regarding the formation scenario \cite{bertoneDarkMatterMounds2024}. Specifically, they consider the formation and growth of a BH following the evolution of a supermassive star positioned at the centre of a DM halo.

DM spikes are not only expected around SMBHs, but also around intermediate mass BHs (IMBH) \citep{zhaoDarkMinihalosIntermediate2005}. 
Studies of such systems using \nbody simulations have been conducted before \citep[e.g.][]{kavanaghSharpeningDarkMatter2024,kavanaghDetectingDarkMatter2020,mukherjeeExaminingEffectsDark2024}. 
These authors set up the DM spike in equilibrium, and they study the effects of the spike profile on inspiralling secondary BHs.
In particular, they find that the effect of dynamical friction \citep[e.g.][]{binneyGalacticDynamicsSecond2008} to play a significant role in speeding up the BH merger. 
Similarly, semi-analytic studies have also shown that dynamical friction can have an impact on the evolution of the binary by making the orbits less eccentric over time \citep[e.g.][]{beckerCircularizationEccentrificationIntermediate2022}.

Non-negligible DM self-interactions will influence the dynamics within a DM spike and lead to some differences with respect to the collisionless case \citep[e.g.][]{alonso-alvarezSelfinteractingDarkMatter2024,fischerDynamicalFrictionSelfinteracting2024, chen2025probingselfinteractingdarkmatter}. 
Assuming velocity-isotropy and spatial-isotropy, \cite{shapiroSelfinteractingDarkMatter2014} solve the gravothermal fluid equations to find a static solution for an SIDM spike profile. 
It was shown that for self-interaction cross-sections with a velocity dependence that scales as $\sigma(v) \propto v^{-a}$,  a stable spike profile exists with a spike index $\gamma = (3+a)/4$. 
The steady-state solutions for SIDM have the same power-law index with and without relativistic corrections. To date, such an 
SIDM system has been studied in the fluid approximation 
but not yet with \nbody simulations \cite{shapiroStarClustersSelfinteracting2018}. 

Observationally, DM spikes are interesting for various reasons. In 2021, NANOGrav \citep{agazieNANOGrav15Yr2023} detected a gravitational wave (GW) signal in the nHz-frequency regime using Pulsar Timing Array (PTA) data.  The leading candidate for the
production mechanism for this signal is the merger of SMBHs.  It was pointed out
that the observations made by PTAs are slightly different from the predictions that arise from the simplest
model of a binary SMBH merger \citep{ellisWhatSourcePTA2024}. {While not statistically significant at this point, it is still interesting to speculate about possible mechanisms that could lead to such a deviation. There are two possible sources:} Either the environment of SMBHs, for example DM spikes, {may have an effect on the merging process leading
to a slightly different spectrum of GWs, or the signal could arise from some other, unrelated, beyond the standard model (BSM) physics}. {See \cite{NANOGrav:2023hvm,ellisWhatSourcePTA2024,burke-spolaorAstrophysicsNanohertzGravitational2019} for an overview of different models.}

Moreover, the SMBH interpretation of the NANOGrav signal is further challenged by the final-parsec problem
\citep{milosavljevicLongTermEvolutionMassive2003}.\footnote{{There is also the interesting possibility that the merging BHs are of primordial origin \cite{Depta:2023uhy}, in which case the final-parsec problem does not arise.}} The simplest models for SMBH mergers assume
that the orbital decay of SMBHs are entirely due to GW energy loss. In this model, the timescale for the SMBHs
to merge when they are separated by a distance of \SI{1}{\pc} is of the order of the Hubble time.
It has been argued that dynamical
friction due to the surrounding DM distribution can accelerate the merger \citep{alonso-alvarezSelfinteractingDarkMatter2024,dosopoulouDynamicalFrictionEvolution2017,kavanaghDetectingDarkMatter2020}, leading to smaller merger timescales, thus effectively 
solving the problem. 

Observations of SMBHs with masses $\gtrsim\SI{1e9}{\solmass}$ at high redshifts are yet to be explained within the framework of \lcdm cosmology \citep{inayoshiAssemblyFirstMassive2020}. 
For such objects to exist, a very massive seed is needed, and various mechanisms have been proposed for the formation of such seeds. 
One such mechanism is based on the gravothermally collapsing nature of SIDM halos \citep{shenMassiveBlackHoles2025,jiangFormationLittleRed2025}. In a similar spirit,
\cite{fengSeedingSupermassiveBlack2021,fengDynamicalInstabilityCollapsed2022} argue that self-interactions can lead to the formation of dense, collapsing cores, which can eventually undergo a gravitational instability leading to the formation of such seeds.

In this paper, we investigate the spike profile as well as the accretion rate of the central SMBH dressed in a DM spike for varying self-interaction cross-sections. To this end, we perform $N$-body simulations. In this work we study velocity-independent self-scatterings and neglect the effects of gas and stars, i.e.\ we concentrate on the behaviour of DM only. 

The structure of the paper is as follows: In \autoref{sec:spike profile}, we briefly introduce DM spikes and introduce equations that give an effective analytic description of these systems. Then, in \autoref{sec:sims}, we describe the setup of our simulations. In \autoref{sec:accretion-cdm} and \autoref{sec:accretion-sidm} we present the results of our study of the accretion rates in DM spikes in CDM and SIDM. Finally, in \autoref{sec:conclusion}, we discuss our results and conclude.
\section{Dark matter spikes}
\label{sec:spike profile}
In this section, we review properties of DM profiles around a BH of mass $\mbh$.
The density profile for DM around a BH is typically parameterised as \citep[e.g.][]{gondoloDarkMatterAnnihilation1999},
\begin{equation}\label{eqn:spike profile}
    \rho(r) = \rhosp \left(\frac{r}{\rsp}\right)^{-\gamma} ,
\end{equation}
where $\rhosp$ and $\rsp$ are the spike density and spike radius, respectively. 
From \cite{gondoloDarkMatterAnnihilation1999,sadeghianDarkmatterDistributionsMassive2013}, it is known that in a fully relativistic analysis, the profile will be suppressed by a factor $F(r) = (1-2R_s/r)^3$, with $R_s$ being the Schwarzschild radius of the central SMBH. 
This factor can be interpreted as depletion of the inner spike at radius $r=2R_s=4\mathrm{G}\mbh/c^2$. 
{As we use Newtonian $N$-body simulations in this work, we neglect this correction in the initial profile. As briefly discussed in the appendix, our implementation should still capture the leading effects.}
The profile in \cref{eqn:spike profile} is valid for, both, CDM and SIDM. 
Given the density profile, the 1D velocity dispersion of the DM spike can be obtained by solving the Jeans equation and is given by
\begin{equation}
    \sod = \dfrac{\mathrm{G}\mbh}{r(1+\gamma)} ,
    \label{eqn:velocity-dispersion-spike}
\end{equation}
assuming the potential is dominated by the BH. For the remainder of the section, we will discuss how the parameters $\rsp, \rhosp$ are determined.
First, we consider the case of CDM with a host halo described by a NFW profile.
For CDM, $\gamma = (9-2\beta)/(4-\beta)$, where $\beta$ is the index with which the DM profile of the host halo scales in the inner region. For example, NFW \citep{navarroStructureColdDark1996}, and Hernquist \citep{hernquistAnalyticalModelSpherical1990} profiles scale as $\rho \propto r^{-1}$ at small radii,
therefore, leading to a spike index of $\gamma=7/3$
\citep{sadeghianDarkmatterDistributionsMassive2013,gondoloDarkMatterAnnihilation1999}. 
We have two more parameters in the spike 
profile, i.e.\ $\rsp$ and $\rhosp$. For a CDM spike, it is conventional 
to define the spike radius as $\rsp:=0.2 r_{\rm in}$, where $r_{\rm in}$ is the 
radius of influence \citep{kavanaghDetectingDarkMatter2020,merrittEvolutionDarkMatter2004,merrittSingleBinaryBlack2004}. The radius of  influence can be determined using
\begin{equation}
    2 \mbh = \int_0^{r_{\rm in}} 4\uppi r^2 \rho(r) \md r ,
    \label{eq:radius-of-influence}
\end{equation}
from which we get the following expression for the spike radius
\begin{equation}
    \rsp=\left[\frac{(3-\gamma)0.2^{3-\gamma}\mbh}{2\uppi\rho_{\mathrm{sp}}}\right]^{1/3} .
    \label{eq:cdm-rsp-equation}
\end{equation}
Another requirement is that the spike profile joins onto the NFW profile of the host halo at $r=\rsp$ \citep{lacroixUniqueProbeDark2017,gorchteinProbingDarkMatter2010}, i.e.,
\begin{equation}
    \rho(\rsp) = \rho_{\rm NFW}(\rsp) .
    \label{eqn:rhosp-cdm-equation}
\end{equation}
\Cref{eq:cdm-rsp-equation,eqn:rhosp-cdm-equation} determine the two parameters $\rsp$ and $\rhosp$ and $\rho_{\rm NFW}(r)$ follows from the properties of the host halo.\\

\begin{figure}
    \centering
    \includegraphics[width=0.45\textwidth]{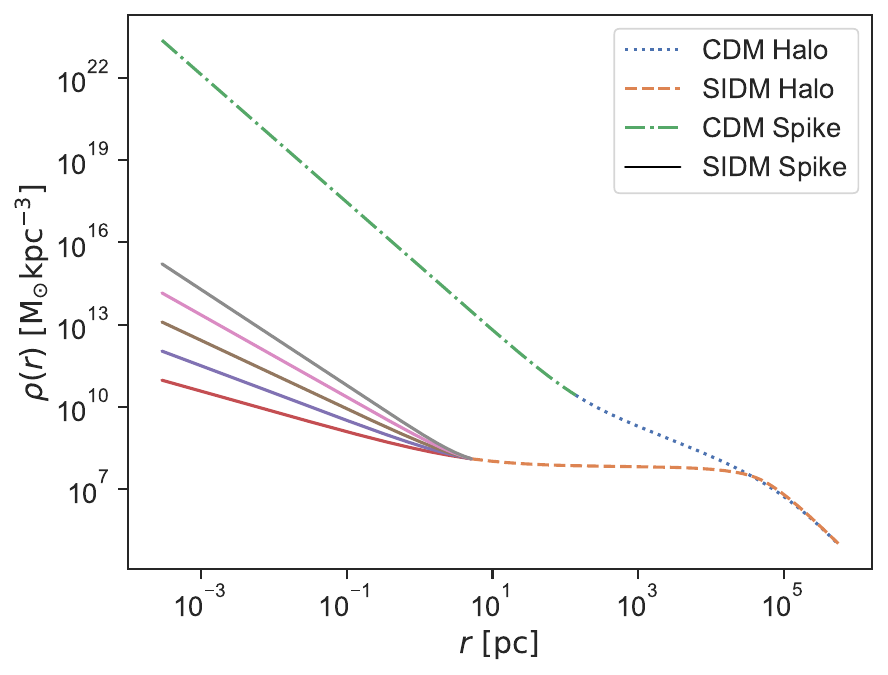}
    \caption{Plot illustrating different DM spike profiles for a SMBH with mass $\mbh=\SI{1e9}{\solmass}$. The corresponding host halo has a mass of $M_{\rm vir} \approx \SI{3e13}{\solmass}$. The solid lines correspond to spike profiles with indices $\gamma = (3+a)/4$, where $a$ is related to the nature of the velocity dependence of the cross-section characterised via $\sigma\propto v^{-a}$. The dashed line corresponds to the SIDM halo of the host halo outside the spike-radius. Dotted lines at the end mark a NFW profile of the host halo in the outer parts. Dashed-dotted lines correspond to the CDM spike with $\gamma=7/3$.}
    \label{fig:spike-illustrative}
\end{figure}
For SIDM, the spike index is set by the nature of the velocity-dependence of the self-interaction cross-section. 
If the velocity dependence is given by $\sigma(v)\propto v^{-a}$, then the spike index is $\gamma = (3+a)/4$, \citep{shapiroSelfinteractingDarkMatter2014}. 

It is worth noting that for $a=4$, the density profile scales as $r^{-7/4}$. The same applies for star clusters, as shown in \cite{bahcallStarDistributionMassive1976}. The similarity of the profiles is due to the Coulombic nature of the interactions, whose cross-section has a $v^{-4}$ dependence. 
The remaining two parameters, $\rsp$ and $\rhosp$, are related to the properties of the host halo, which we assume to have a cored profile \citep{adhikariAstrophysicalTestsDark2022} 
with a core radius $r_c$. Within the core, we have constant velocity dispersion, $\sigma_0$, and a constant core density, $\rho_0$. The host halo is usually assumed to be a NFW profile initially, and over time, the halo develops an isothermal core due to self-interactions, while the outer regions are still well described by a NFW profile. The parameters $\lbrace \sigma_0,\rho_0 \rbrace$ are determined from $\sbm$ and the NFW parameters of the initial halo in two steps: Firstly, the core radius $r_c$ is determined from the scattering cross-section $\sbm$ by demanding that at least one scattering event has taken place given within a given time $t$ (assuming $t=0$ as initial condition). Specifically we have: 
\begin{equation}
    \dfrac{\langle \sigma v_{\rm rel} \rangle}{m_\chi} \rho_{\rm NFW}(r_c(t)) \cdot t \sim 1 ,
    \label{eqn:core-radius}
\end{equation}
where $v_{\rm rel}$ is the relative velocity between DM particles. 
This equation implies that, starting from the initial NFW profile, the core radius will grow with time. Note however that the timescale of spike formation is typically much shorter than the self-interaction relaxation timescale. If the central black hole grows adiabatically, we would therefore expect the formation of a steep CDM spike initially, which will then subsequently relax to the steady state SIDM solution described above (cf also Fig.~\ref{fig:spike-illustrative} for a visualisation of the overall system at late times and for different velocity dependencies of the SIDM scatterings).
If we are interested in the spike properties today, we would take $t=t_\text{age}$ with $t_\text{age}$ the age of the system under consideration. 

To fully determine the SIDM halo parameters, one needs to also solve the spherical Jeans equation $\nabla(\rho_{\rm DM} \sigma_0^2) = -\rho_{\rm DM} \nabla\Phi(r)$ along with the Poisson equation $\nabla^2 \Phi(r) =-4\uppi \mathrm{G} \rho_{\rm DM}$ (see \citep{kaplinghatDarkMatterHalos2016,jiangSemianalyticStudySelfinteracting2023} for more details).
The spike radius and spike density are then given by \citet{shapiroSelfinteractingDarkMatter2014,alonso-alvarezSelfinteractingDarkMatter2024},
\begin{equation}
\rsp = \dfrac{\mathrm{G} \mbh}{\sigma_0^2},
\label{eqn:spike-radius-from-sigma}
\end{equation}
and
\begin{equation}    
\rhosp = \rho_0
\label{eq:spike-radius-sidm} ,
\end{equation}
respectively. We now study the evolution of spikes in the presence of DM self-interactions with numerical simulations, which we will describe in the next section.


\section{Simulation setup}
\label{sec:sims}
We employ the cosmological, hydrodynamical $N$-body code \textsc{OpenGadget3} \citep[e.g.][Dolag et al.\ in prep.]{grothCosmologicalSimulationCode2023,ragagninExploitingSpaceFilling2016} a derivative of \textsc{Gadget-2} \citep{springelCosmologicalSimulationCode2005}
that performs gravitational force calculations using a tree 
method.  

\subsection{Numerical setup}\label{sec: sidm implementation}

We simulate velocity-independent, isotropic self-interactions.
The self-interactions are modelled in a Monte-Carlo fashion by scattering pairs of numerical particles using \textsc{OpenGadget3}. The particular implementation used is described as rare self-interactions in \cite{fischerNbodySimulationsDark2021,fischerCosmologicalSimulationsRare2022, fischerCosmologicalIdealizedSimulations2024}. 
In this scheme, the probability that a numerical particle labelled $i$  with mass $m_i$ scatters off another numerical particle $j$ with mass $m_j$ is given by
\begin{equation}\label{eqn : scattering probability}
    P_{ij} = \frac{\sigma(v_{ij})}{m_\chi} \ m_j \ |v_{ij}| \ \Delta t  \ \Lambda_{ij} ,
\end{equation}
where $v_{ij}$ is the relative velocity between the numerical particles $i$ and $j$, $\Delta t$ 
is the time-step used in the simulation, $\Lambda_{ij}$ is the kernel overlap integral 
and $\sigma(v_{ij})/m_\chi$ is the total cross-section 
per unit mass of DM particle. 
Note that the indices $i,j$ are not summed over. Each particle is assigned a kernel adaptively such that each particle has at least $N_{\rm ngb}$ number of neighbours. In addition, all numerical DM particles have the same mass, i.e.\ $m_i=m_j=M_{N,DM}$.
For more details on the implementation and the choice for the kernel, see 
appendix B of \cite{fischerNbodySimulationsDark2021}. A collection of kernels used in 
other modern implementations of SIDM in \nbody codes can be found in
\citet[equations~11--15]{adhikariAstrophysicalTestsDark2022}. The SIDM implementation, we use, explicitly conserves energy and linear momentum \citep{fischerNumericalChallengesEnergy2024}.

We model accretion the following way: at the end of each time-step, we remove every particle that passes the innermost stable circular orbit {$\risco = 3 R_S = 6{\rm G}\mbh/c^2$, of the (assumed to be non-rotating)} central BH. Here $c$ is the speed of light. 
{As mass accretion is negligible in all our simulations, i.e. $\Delta \mbh / \mbh \lll 1$, it is a good approximation to assume that the total BH mass does not change.} 
The accretion rate is then given by the loss rate of the total number of particles.

\subsection{Precision parameters}

The simulations contain a number of parameters that control the precision of the force calculations.
We fix the gravitational softening length{, $\epsilon_\mathrm{soft}$,} for the particles to be approximately the Schwarzschild radius 
of the BH, and it sets the smallest length scale of the simulated system. The opening angle for the gravitational force calculations using the tree method is chosen to be small enough  {to ensure that errors arising from the asymmetric force evaluation of the one-sided oct-tree \citep{barnesHierarchicalLogForcecalculation1986} are negligible}.
Similar values have been adopted by \cite{kavanaghDetectingDarkMatter2020}. We provide the numerical values in \cref{tab:precision}. 

For all simulations we fix the time-step, $\Delta t$, to a constant value which is the same for all 
particles. 
We denote the maximum allowed value of the gravitational time-step corresponding to the gravitational force calculation as 
$\Delta t_{\rm grav} $. This is in practice at least 100 times smaller than the smallest dynamical timescale in the system. 
For simulations with SIDM, there is an additional condition that we have to respect for choosing the  time-step, $\Delta t_{\rm SIDM}$. The condition being that the time-step must be small enough to ensure that the scattering probability stays well below unity. For more information see \cite{fischerCosmologicalIdealizedSimulations2024}. 
Therefore, we choose the minimum of the two, i.e.\ $\Delta t = {\rm min}(\Delta t_{\rm grav} , \Delta t_{\rm SIDM})$. We ensure that this criterion is fulfilled for the entire simulation period. In \cref{sec:energy-conservation}, we show that energy is conserved numerically to within 0.1\%. The time step is chosen to be small enough such that this criterion is fulfilled during the entire simulation.

\begin{table}[]
\caption{Precision parameters used in \textsc{OpenGadget3}.}
\centering
\begin{tabular}{|c|c|}
\hline
Parameter & Value \\ \hline
\texttt{ErrTolForAcc}        & \num{1e-4} \\
${\epsilon_{\rm soft}}$ & \SI{0.0017}{\pc} \\
$\Delta t$ & $10^{-4}$ yr\\
$N_{\rm ngb}$ & 48\\ \hline
\end{tabular}
\label{tab:precision}
\end{table}

\begin{figure}
    \centering
    \includegraphics[width=0.45\textwidth]{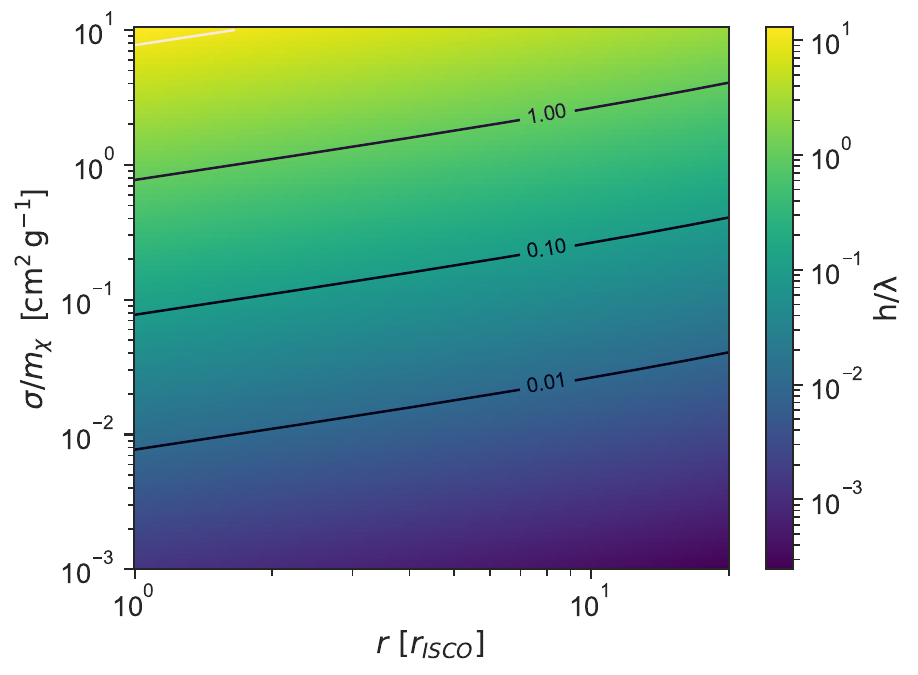}
    \caption{Two-dimensional histogram of the kernel size to mean-free-path ratio as a function of radius (in units of $\risco$) and  cross-section, $\sigma/m$. The black lines correspond to constant values of $h/\lambda$.}
    \label{fig:kernel-size-ratio}
\end{figure}

As described in the \cref{sec: sidm implementation}, the scattering probability is calculated using kernels. 
The size of the kernel is chosen such that there are $N_{\rm ngb}$ number of particles within the kernel. 
The kernel size must be smaller or at the most be {roughly} equal to the physical mean-free-path of SIDM scatterings. 
This requirement ensures that, energy and momentum are not transferred per scattering on length scales that are greater than the mean-free-path. The ratio between the kernel size, $h$, and the mean free path, $\lambda$,
\begin{equation}
    \label{eqn:kernel size}
\frac{h}{\lambda} \approx \frac{\sigma}{m_\chi} \, \rho^{2/3} \sqrt[3]{\frac{3\,M_{\rm N,DM}\,N_\mathrm{ngb}}{\uppi \sqrt{2}}}  ,
\end{equation}
therefore indicates which cross-sections can be simulated faithfully.
This ratio should at most be of order of unity in order to yield reliable results \citep{fischerNumericalChallengesEnergy2024}.
See \cref{fig:kernel-size-ratio} for a two-dimensional histogram of the kernel size to mean-free-path ratio as a function of radius (in units of $\risco$) and  cross-section, $\sigma/m$. This figure shows that for radii close to the ISCO radius and $\sbm = \SI{0.8}{\cm\squared\per\g}$, the ratio $h/\lambda$ gets larger than unity. In all other parts of this parameter space, this value is smaller than 1. 

\subsection{Astrophysical parameters}\label{sec:astrophysical-parameters}

To ensure that we have a sufficient number of particles in the inner region to faithfully simulate the interesting physics of the spike, we adopt a slightly different prescription of the initial condition in our numerical analysis. Specifically we simulate the DM density around the BH with a truncated (so-called $\alpha\beta\gamma$) spike profile, which is given by
\begin{equation}
    \rho(r) = \rhosp \left(\dfrac{r}{\rsp}\right)^{-\gamma}  \left(1+\left(\dfrac{r}{r_t}\right)^{\beta}\right)^{-\alpha}, 
    \label{eqn:truncated-spike-profile}
\end{equation}
where $\rsp$ is the spike radius and $\rhosp$ the spike density, as before. Rather than imposing a cut `by hand' in \cref{eqn:spike profile},
the profile is modified at large radii by a power law with a steeper slope, i.e., $\alpha > \gamma$ according to \cref{eqn:truncated-spike-profile}, which provides a smooth truncation to the profile for larger radii. 
This truncated profile has been employed previously for the case $\beta=1$ in the context of CDM \nbody simulations in \cite{kavanaghDetectingDarkMatter2020}.

We sample the radii using the spike profile as the target distribution, and for sampling velocities we use the
Eddington-inversion method 
\citep{binneyGalacticDynamicsSecond2008,lacroixAnatomyEddingtonlikeInversion2018}. 
In our simulation we sample particles in the radial range between $\rmin = \risco$ 
and $\rmax= 0.1\rsp$.
We simulate a system that is similar to the OJ287 system described in
\cite{deyAuthenticatingPresenceRelativistic2018,alachkarDarkMatterConstraints2023}, which has a binary SMBH at the centre. In \cite{deyAuthenticatingPresenceRelativistic2018} they estimate the mass of the primary SMBH $M_{\rm BH}$ to be \SI{1.8e10}{\solmass} and the secondary SMBH to be \SI{1.5e8}{\solmass}. Since we are interested in studying an isolated DM spike, we only consider the primary SMBH in this work.
The NFW profile parameters of the corresponding host halo follows from \cite{chanFirstRobustEvidence2024}.
The virial mass of the host galaxy of OJ287 is obtained to be $M_{\rm vir} \approx \SI{2e14}{\solmass}$ from the mass 
of the SMBH. This is achieved using the scaling relations between the virial mass of the host and the mass of the central SMBH \citep{bandaraRelationshipSupermassiveBlack2009a}.  Then, given the virial mass, the NFW parameters can be determined using  mass-concentration relations \citep{duttonColdDarkMatter2014}, yielding a concentration of $c\approx4.8$.

In SIDM, for any given self-interaction cross-section, we solve the Jeans-Poisson equations to determine $\lbrace \rho_0, \sigma_0 \rbrace$ of the SIDM halo. To this end, we use the publicly available isothermal code provided in \cite{jiangSemianalyticStudySelfinteracting2023}. 
The only input needed is the time the system had to evolve to set the initial condition. In the following, we use the approximate age of the system ($t_{\rm age}$) to solve \cref{eqn:core-radius}, noting that at earlier times the spike density will have been larger. 
We find that for a scattering cross-section of $\sbm = \SI{0.1}{\cm\squared\per\gram}$ and an age of $t_{\rm age} = \SI{10}{\giga\yr}$ the resulting density is
$\rho_0 \sim \SI{10}{\solmass\per\pc\cubed}$. The choice of these parameters should not  affect our results that we present in the next sections.

\begin{table}
    \caption{Physical parameters of the DM spike.}
    \centering
    \renewcommand{\arraystretch}{1.5}
    \label{tab:cdm-7b3-oj287}
    \begin{tabular}{|c|c|c|c|c|c|}
        \hline
        $\rho_{\rm sp}$ & $\gamma$ & $r_{\rm sp}$ & $\alpha$ &  $r_{\rm t}$ & $\sbm$ \\
        (\unit{\solmass\per\pc\cubed}) & & (\unit{\kpc}) &  & ($\risco$) & (\unit{\cm\squared\per\g}) \\ 
        \hline
        20 & 3/4 & 1.5 & 4 &  200 & {0}  \\
        \hline
        20 & 3/4 & 1.5 & 4 & 200 & {0.2} \\
        \hline
        20 & 3/4 & 1.5 & 4 & 200 & {0.4} \\
        \hline
        20 & 3/4 & 1.5 & 4 & 200 & {0.6} \\
        \hline
        20 & 3/4 & 1.5 & 4 & 200 & {0.8} \\
        \hline
        20 & 3/4 & 1.5 & 4 & 200 & {1.4} \\
        \hline
        20 & 3/4 & 1.5 & 4 & 200 & {2.0} \\
        \hline
        20 & 3/4 & 1.5 & 4 & 200 & {4.0} \\
        \hline
        20 & 3/4 & 1.5 & 4 & 200 & {8.0} \\
        \hline
        20 & 1/2 & 1.5 & 4 & 500 & {\num[exponent-product=\cdot]{1.0}}\\
        \hline    
        20 & 7/4 & 1.5 & 2 & 200 & {\num[exponent-product=\cdot]{2e-5}} \\
        \hline
    \end{tabular}
    \tablefoot{The columns from the left to the right are the following: spike density $\rhosp$, spike index $\gamma$, spike radius $\rsp$, truncation index $\alpha$, truncation radius $r_t$, and finally the cross-section per unit mass $\sbm$. The first row corresponds to the collisionless case, which we nevertheless assume to have the same initial spike profile to enable a more direct comparison. 
    For all our simulations, we have set the profile parameter $\beta$ to be 1. 
    }
\end{table}

The gravitational timescale is simply given by
\begin{equation}
    t_{\rm G} \sim r^{3/2} (\mathrm{G} M_{\rm BH})^{-1},
\end{equation}
since $M_{\rm BH}$ contributes the most to the gravitational dynamics. 

Finally, the relaxation timescale due to self-interactions is given by
\begin{equation}
    t_{\rm SIDM} = \dfrac{1}{\rho(r) \sbm \sod} \simeq    \frac{r^{\gamma+1/2}}{\sqrt{\mathrm{G} \mbh} \rsp^{\gamma} \rhosp \sbm},
    \label{eqn:relaxation timescale}
\end{equation}
{where in the last step we used} $\sod^2(r) \sim \mathrm{G} \mbh/r$ from \cref{eqn:velocity-dispersion-spike} and the density profile given in \cref{eqn:spike profile}.
\section{Black hole mass accretion with CDM}
\label{sec:accretion-cdm}

In the case of collisionless DM, there are no processes
that can lead to the loss of angular momentum, which prevents the effective accretion of DM particles. 
Still, there can be some CDM particles with insufficient angular momentum and energy such that they get accreted by the BH. 
In this section {-- in addition to the numerical analysis --  we attempt an order-of-magnitude analytical estimate} for the accretion rate given the initial conditions. 
A fully relativistic counterpart of this calculation can be found in \cite{shapiroSpikesAccretionUnbound2023}. 
Since the major contributor to the gravitational potential is the SMBH, the accretion rate of CDM through a spherical surface can be calculated from the initial phase-space distribution, $f(E,J)$, where $E$ and $J$ are energy and angular momentum, respectively. Here, $f(E,J)$ is normalised to yield the total mass when integrated over the full phase space, i.e.\
\begin{equation}
    \rho(r) = \int f(E,J) \dd^3 \mathbf{v} .
\end{equation}
In CDM spikes, the contribution to the accretion comes from particles that initially have a mostly radial velocity with small angular momentum. The mass of the phase space patch with the radial velocity directed towards the BH is given by
\begin{equation}
    \dd M(\mathbf{x},\mathbf{v}) = \dfrac{1}{2} f(\mathbf{x},\mathbf{v})\, \dd^3 \mathbf{x}\, \dd^3 \mathbf{v} ,
\end{equation}
where the factor 1/2 accounts for the radially inward moving particles. Since the initial distribution is spherically symmetric and isotropic in, both, coordinate and velocity subspaces, we can write for the mass in the interval $\dd r\,\dd E\,\dd J$ centred on $(r,E,J)$:
\begin{equation}    
       M^{-}(r,E,J)\, \dd r\, \dd E \, \dd J = 8\uppi^2  f(E,J) \dfrac{J}{|v_r|} \dd r\, \dd E\, \dd J , 
\end{equation}
{where the "$^-$" indicates that it is the mass that is moving inwards and will be accreted.}
Hence, the accretion rate through a spherical surface of radius $R$ is given by
\begin{align}
    \dot{M} &= \int_{\Phi(R)}^{\infty} \dd E \int_0^{J_{\rm min}} \dd J\, |v_r| M^{-}(r,E,J)\\
            &= \int_{\Phi(R)}^{\infty} \dd E \int_0^{J_{\rm min}} \dd J\,  |v_r| 8\pi^2  f(E,J) \dfrac{J}{|v_r|} \\
    &=8\uppi^2 \int_{\Phi(R)}^\infty \dd E \int_0^{J_{\rm min}} f(E,J)\, J\dd J .\label{eqn:cdm accretion rate}
\end{align}
For our system, the distribution function has no dependence on angular momentum, and therefore the integral can be evaluated numerically using the $f(E)$ we constructed using the Eddington inversion method. 

When computing the integral we set the upper limit for the energy integral to 0 instead of $\infty$ since the spike is gravitationally  bound and has been set up dynamically stable in the ICs. Hence, no particle has an energy greater than 0 by construction. 
Similarly, the lower limit for the energy integral cannot be smaller than the potential energy on the surface of interest. 

The minimum angular momentum for a bound orbit is, $J_{\rm min} = R \sqrt{2\left(E+\mathrm{G}M/R\right)}$, and this expression follows from the Keplerian two-body energy of the DM-SMBH system evaluated at the pericentre, with the pericentre being at the imaginary radial surface. 

{Taking the same halo parameters as used for the simulation (cf.~table~\ref{tab:precision}), the integral above evaluates to $\SI{5.1}{\solmass\per\yr}$ for $R=\risco$. In the numerical simulations we find a somewhat smaller value, $ \dot{M}_{\rm sim} \approx \SI{1.7}{\solmass\per\yr}$, which we estimated} by linear fitting for an initial period of {\SI{15}{\yr}}, where the accretion rate is approximately linear. For later stages, due to the loss cone depletion and the fact that the lost mass is not replenished from the outer halo, {the accretion rate will decrease further}. 
{This implies that $f(E,J)$ of the simulated system will slowly change in the CDM case due to accretion. In contrast, in the SIDM case, $f(E,J)$ is affected by the DM scattering events even in the absence of accretion.} 

\begin{figure}
    \centering
    \includegraphics[width=0.49\textwidth]{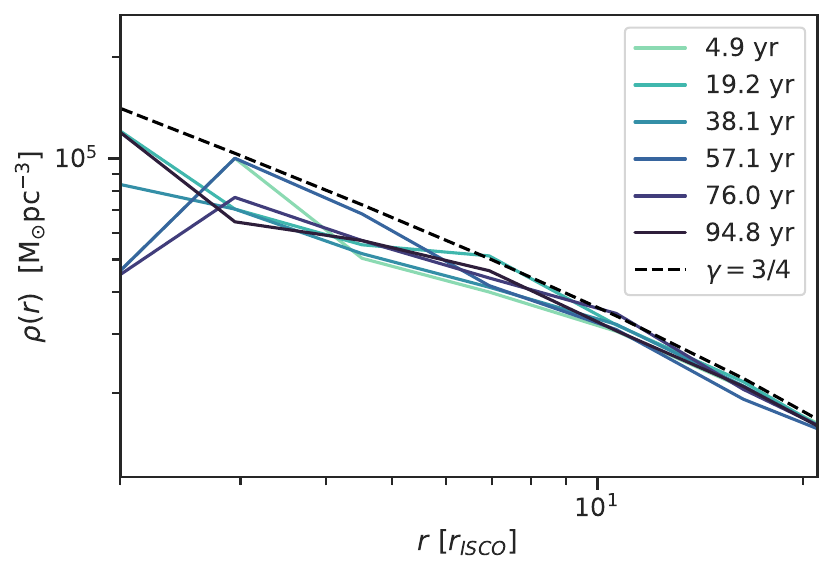}
    \caption{Evolution of the CDM spike profile. The initial condition contains particles sampled from the profile given by the dashed line. The bottom panel displays the deviation from the initial profile.}
    \label{fig:cdm-g3b4-density-profile}
\end{figure}
{The result of our \nbody simulation for the collisionless case is shown in \cref{fig:cdm-g3b4-density-profile}.} The profile undergoes evolution due to accretion. 
As time progresses, the accretion rate goes down as the loss cone is getting depopulated. This leads to a stable profile at later times. 
{Note that in our simulation we start from a fully populated spike with parameters as given in table~\ref{tab:precision} and do not take into account the spike formation.}
In reality, as the CDM spike forms due to adiabatic contraction, it is natural to expect that the particles in the loss cone are steadily accreted during the formation of the spike. Since we start with a profile that has not undergone such a process, {the initial accretion is artificially high in the CDM simulations.}

\section{Black hole mass accretion with SIDM}
\label{sec:accretion-sidm}

{Let us start this section with a discussion of the gravothermal fluid equations in order to obtain a qualitative understanding of the behaviour of the spike index as well as the} accretion rates for varying cross-sections.
{The gravothermal fluid equations can be written as}
\citep{yangGravothermalEvolutionDark2022,balbergSelfInteractingDarkMatter2002,kodaGravothermalCollapseIsolated2011a,essigConstrainingDissipativeDark2019,shapiroSelfinteractingDarkMatter2014}:
\begin{align}
    \frac{\partial M}{\partial r}&=4\uppi r^{2}\rho, \label{eqn:mass_conservation} \\ 
    \frac{\partial(\rho \sod^2)}{\partial r}&=-\frac{\mathrm{G} M\rho}{r^2},\label{eqn:hydrostatic_equilibrium} \\
    \frac{\partial L}{\partial r} &= 4\uppi r^2 \rho \sod^2 \frac{D}{D t}\ln\frac{\sod^2}{\rho} \label{eqn:thermodynamical-law}, \\ 
    \frac{L}{4\uppi r^{2}}&=-m_\chi\kappa\frac{\partial \sod^2}{\partial r}. \label{eqn:heat_flux} 
\end{align}
Here, $D/Dt$ denotes a Lagrangian time derivative, $\kappa$ characterizes the heat conductivity, $L$ 
is the luminosity, and $M$ is the enclosed mass. The luminosity quantifies the 
heat flux arising due to self-interactions. A positive luminosity implies that energy is flowing outwards.  
\Cref{eqn:mass_conservation,eqn:hydrostatic_equilibrium,eqn:thermodynamical-law}
are the first three moments of the collisional Boltzmann equation and \cref{eqn:heat_flux} provides a closure relation.

\Cref{eqn:mass_conservation} is the continuity equation corresponding to the conservation 
of mass, \cref{eqn:hydrostatic_equilibrium}  is the equation of hydrostatic equilibrium,
\cref{eqn:thermodynamical-law} corresponds to the first law of thermodynamics. 
Finally, \cref{eqn:heat_flux} defines the heat flux. 
\begin{figure}
    \centering
    \includegraphics[width=0.45\textwidth]{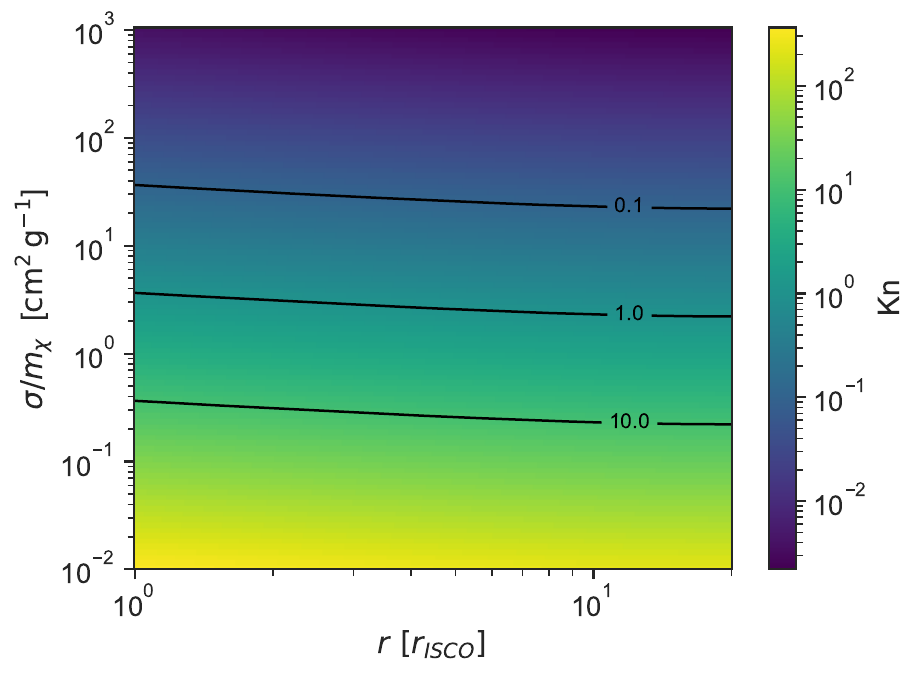}
    \caption{Radial profile of Knudsen number for varying cross-sections. 
   }
    \label{fig:knudsen-number}
\end{figure}

If the evolution is dominated by the effects of SIDM (instead of gravity), the system is said to be in the short-mean-free-path (SMFP) regime, while if the evolution is dominated by effects of gravity the system is said to be in the long-mean-free-path (LMFP) regime.  
These regimes are delineated with the help of the Knudsen number. It is defined as the ratio of the scale height to mean-free-path corresponding to SIDM scatterings. 
\begin{equation}
    {\rm Kn}(r) = \dfrac{\lambda(r)}{H(r)}= \dfrac{1}{r \rho(r) \sbm} .
    \label{eqn:knudsen-number}
\end{equation}
Here, $\lambda$ is the mean free path for a given self-interaction cross-section, i.e.\ $\lambda = m_\chi/(\rho \sigma)$. 
$H$ is the gravitational scale height, {effectively given by the distance from the BH, $H \simeq t_G \sigma_{\rm 1D} \sim r^{3/2} r^{-1/2} = r$}~\citep{shapiroSelfinteractingDarkMatter2014}. 
We show the radial profile of the Knudsen number {for the case $\gamma = 3/4$ in \cref{fig:knudsen-number}, observing a scaling $\sim r^{-1/4}$ as expected}.
In the LMFP, $\mathrm{Kn}\gg1$ and in the SMFP, $\mathrm{Kn} \ll 1$. We further identify $\mathrm{Kn}\sim1$ as the intermediate-mean-free-path (IMFP) regime. 
Depending on the regime, the nature of thermal conductivity changes. Under the assumption of velocity isotropy, the conductivity in these regimes scales as
\begin{equation}
\kappa_{\rm lmfp} \propto \rho \frac{H^2}{t_{\rm SIDM}} 
\quad \text{yielding}  \quad 
L_{\rm lmfp} \propto \frac{1}{t_{\rm SIDM}} \propto (\sbm)\label{eq:conductivity-lmfp}
\end{equation}
for the LMFP and
\begin{equation}
     \kappa_{\rm smfp} \propto \rho \frac{\lambda^2}{t_{\rm SIDM}} \quad \text{yielding}  \quad L_{\rm smfp} \propto \frac{\lambda^2}{t_{\rm SIDM}} \propto (\sbm)^{-1} \\\label{eq:conductivity-smfp}
\end{equation}
for the SMFP.

In the weakly collisional limit, $\lambda\gg H$, and the system is in the LMFP. On the other hand, 
in the limit of large cross-section, the system will be in the SMFP regime, 
since $\lambda \ll H$. Moreover, for $\sbm\rightarrow\infty \implies \kappa_{\rm smfp} \rightarrow 0$, implying that the SIDM effectively behaves like a non-conducting fluid {in this limit}.
Given the conductive nature of SIDM spikes, one can seek a steady-state solution to the spike density profile, by demanding that the spike has a constant luminosity. Such a constant luminosity could lead to a steady-state accretion \citep{pringleAccretionDiscsAstrophysics1981,frankAccretionPowerAstrophysics2002,carrollIntroductionModernAstrophysics2017}.
This requirement can be used to determine what the spike-index needs to be. For example, in the constant cross-section case, we require from \cref{eqn:heat_flux,eq:conductivity-lmfp} that $L\propto r^{3/2} r^{-2\gamma} = \mathrm{const.} $, and this results in $\gamma=3/4$.

\begin{figure*}[t]
    \includegraphics[width=0.45\linewidth]{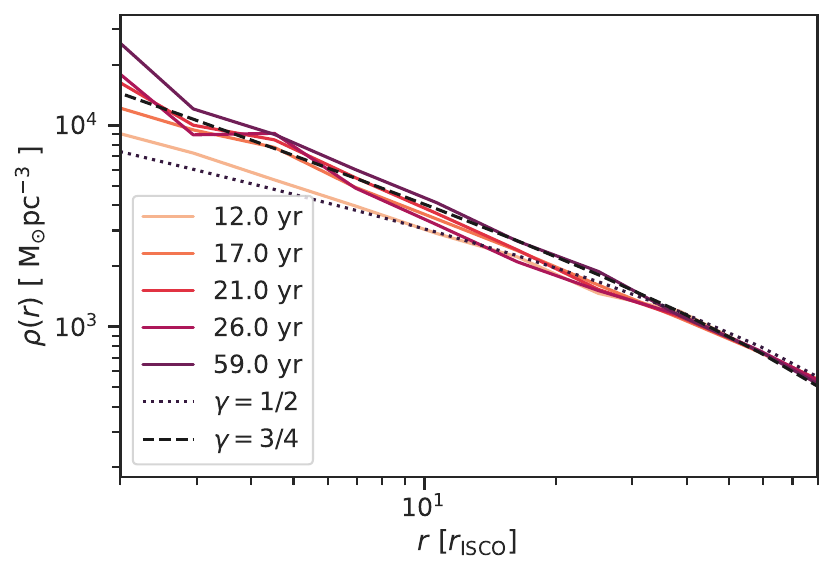}
    \includegraphics[width=0.45\linewidth]{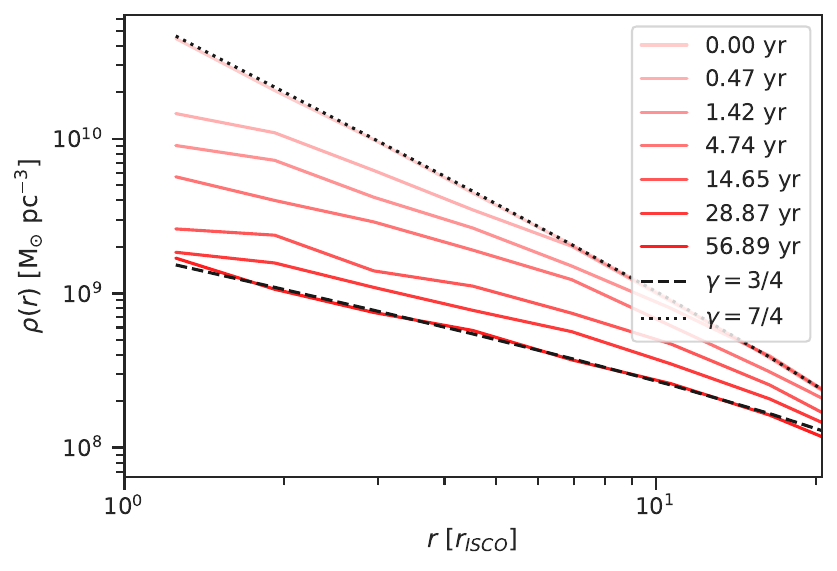}
    \caption{
    Spike density profile in the inner regions at various times for a DM spike setup initially with two different spike indices. Left and right panel correspond to $\lbrace\gamma,\sbm\rbrace$ of  $\lbrace 1/2,\SI[exponent-product = \cdot]{1.0}{\cm\squared\per\g}\rbrace$, and $\lbrace 7/4,\SI[exponent-product = \cdot]{2e-5}{\cm\squared\per\g}\rbrace$ respectively. In both panels, the dotted lines are the initial density profile that is sampled and dashed lines correspond to a spike profile with index $3/4$.  For both steeper and shallower starting profiles we see that the spike index evolves towards $\gamma=3/4$.}
    \label{fig:varying_gamma}
\end{figure*}

Let us now present the numerical evolution of the spike profile and compare it to analytical expectations. Specifically, we are interested in the question if and how quickly the steady-state profiles are reached if we start with profiles that are either steeper or shallower than the expected equilibrium state with $\gamma=3/4$. In \cref{fig:varying_gamma} we show the evolution for initial values $\gamma=7/4$ and $\gamma=1/2$, respectively. 
We immediately see that for both cases the system transitions quickly towards the steady-state solution, $\gamma=3/4$. 
This result is expected because, for $\gamma\neq3/4$ and for velocity-independent scatterings, the luminosity is not constant in space nor time. 
Effectively, this leads to accretion rates that vary with time until the system settles into the static profile.
We also see that the spike radius of the newly forming profile is growing with time. 
Note that we have used different cross-sections in the simulations owing to the different densities reached in the inner regions (cf.~table~\ref{tab:cdm-7b3-oj287}).
{We expect that also for larger radii and on larger timescales the mass that the DM spike loses to the BH gets replenished. 
It will be interesting to test this in simulations in a cosmological setting.}

The steady-state spike solutions introduced in \cite{shapiroSelfinteractingDarkMatter2014} are valid in the LMFP regime. In the other extreme, we have the hydrodynamic limit, where the static solution requires a spike index of $3/2$ \cite{shapiroSelfinteractingDarkMatter2014}. 
In our simulations we however concentrate on the LMFP and IMFP regimes. 

\begin{figure}
    \centering
    \includegraphics[width=0.45\textwidth]{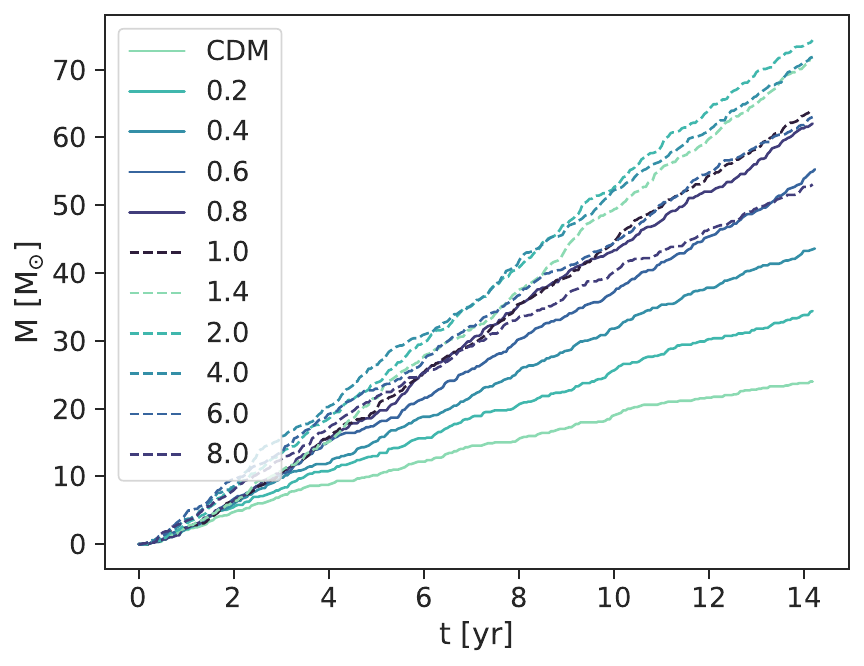}
    \caption{Accreted mass vs. time for varying cross-sections. The labels indicate the cross-section in units of \unit{\cm\squared\per\g}.
    Solid lines correspond to $\sbm<\SI{1.0}{\cm\squared\per\g}$ and dashed lines to $\sbm \geq \SI{1.0}{\cm\squared\per\g}$.
    }
    \label{fig:accretion-infall-vs-time}
\end{figure}

Let us now turn to the discussion of the accretion onto the central BH.
In Fig.~\ref{fig:accretion-infall-vs-time} we plot the accreted mass versus time for a range of cross-sections including the collisionless case for comparison. As expected, the SIDM accretion rates are larger than the CDM accretion rates for the same initial conditions. 
We indicate cross-sections $\sbm<\SI{1}{\cm\squared\per\g}$ by solid lines, and $\sbm \ge \SI{1}{\cm\squared\per\g}$ by dashed lines,
observing that the kernel-size to mean-free-path ratio becomes greater than 1 for cross-sections $\gtrsim \SI{1}{\cm\squared\per\g}$. 

\begin{figure}
    \centering
    \includegraphics[width=0.45\textwidth]{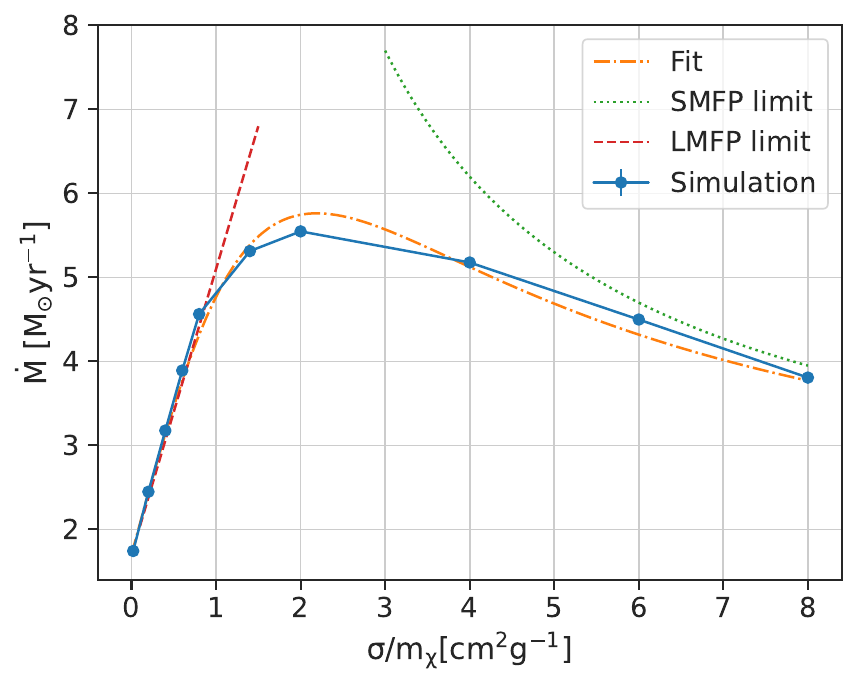}
    \caption{Initial accretion rate vs.\ $\sbm$ for all simulated cross-sections. The solid line corresponds to the accretion rate inferred from the simulation, dashed-dotted line to the full fit, dashed line to the LMFP limit of \cref{eqn:imfp-fit}, dotted line to the SMFP limit of \cref{eqn:imfp-fit}.}
    \label{fig:imfp-fit}
\end{figure}

{In \cref{fig:imfp-fit} we show the simulated accretion rates as a function of the self-interaction cross-section. We observe that within the LMFP regime the accretion rate increases linearly to an excellent approximation.
However, for cross-sections $\sbm\gtrsim\SI{1.0}{\cm\squared\per\g}$, the mass accretion rate does no longer grow linearly but increases more slowly and for even larger cross-section starts to decrease.  This qualitative trend can be understood as being due to the transition from the LMFP regime to the SMFP regime for increasing cross-sections. The fact that cross-sections of the order \SI{1}{\cm\squared\per\g} correspond to the transition regime can also be inferred from \cref{fig:knudsen-number}.}

In the following let us try to obtain an analytical understanding of the observed accretion rates.
We have already established that the accretion rates are proportional to the luminosity $L$ in the gravothermal fluid equations and the different scaling in the LMFP and SMFP regimes, cf. \cref{eq:conductivity-lmfp} and \cref{eq:conductivity-smfp}.
We therefore make the following ansatz for the fitting function,
\begin{equation}
    \dot{M} = C + \left(A \sbm + \dfrac{B}{\sbm}\right)^{-1} .
    \label{eqn:imfp-fit}
\end{equation}
The RHS of the above equation is also motivated by the standard expression for thermal conductivity in the intermediate-mean-free-path (IMFP) regime \citep{yangGravothermalEvolutionDark2022,kodaGravothermalCollapseIsolated2011a}. The expression is a simple smooth interpolation from the LMFP regime that has $\kappa_{\rm lmfp} \propto \sbm$ to the SMFP regime with $\kappa_{\rm smfp}\propto(\sbm)^{-1}$. Alternate interpolating functions have also been discussed in the literature \citep{maceCalibratingSIDMGravothermal2025,nishikawaAcceleratedCoreCollapse2020}.
In \cref{fig:imfp-fit}, we show accretion rates extracted from the simulation, along with \cref{eqn:imfp-fit} fitted with the simulation data. 
The fitted values are: $A\approx\SI{0.056}{\per\solmass\yr\per\cm\squared\g} \approx 585 \rsp^{-3/2}\rhosp^{-1}(\mathrm{G}\mbh)^{-1/2}\unit{\per\cm\squared\g}$, $ B\approx\SI{0.27}{\per\solmass\yr\cm\squared\per\g} \approx 2823 \rsp^{-3/2}\rhosp^{-1}(\mathrm{G}\mbh)^{-1/2}\unit{\cm\squared\per\g} $ and $C \approx \SI{3.67}{\solmass\per\yr\per\cm\squared\g} \approx \num{1.6e-4}\rsp^{3/2} \rhosp (\mathrm{G}\mbh)^{1/2}$, where in the second step we expressed the constants in units comprised of the parameters of the system.
To extract the LMPF limit from the above, we expand the RHS around zero and retain only the lowest-order contribution. Thus, we get, $\dot{M} \approx C + B^{-1} \sbm $. 
The constant factor $C$ captures the CDM limit, that is it captures the contribution to the accretion rate coming from the particles that have $J<J_{\rm min}$. The second term directly captures the contribution to the accretion rate coming from the self-interactions. In appendix~\ref{app:accretion} we provide an analytical evaluation of this coefficient.
Similarly, for the SMFP limit, we expand the RHS around $\infty$, and we obtain $\dot{M}\approx C + A (\sbm)^{-1}$. 
We show these fits along with the data points in \cref{fig:imfp-fit}, observing an
excellent agreement with our numerical results.

\subsection{Maximum accretion rate}
 In this  subsection we extend the validated model to DM spikes with different spike parameters and SMBH masses and calculate the corresponding accretion rates. Then, we compare it to the Eddington accretion rate $\medd$. The Eddington accretion rate, $\medd$, depends linearly on $\mbh$, i.e. $\medd = L_{\rm Edd} (0.1 c^2)^{-1} \approx (\num{2.3e-8} \unit{\solmass\per\yr}) \mbh/\unit{\solmass}$ \citep{inayoshiAssemblyFirstMassive2020}.
 
 In the simulations we have run, we see in \cref{fig:imfp-fit} that the accretion rate maxes out around $\SI{2.0}{\cm\squared\per\g}$ in the IMFP regime. This regime can be identified with $\rm Kn(\risco)\sim1$. Therefore,  we { approximate} the maximum accretion rate as the accretion rate corresponding to $\rm Kn(\risco)=1$. For a SMBH-spike system characterized by the parameters $\lbrace\rsp,\rhosp,\mbh\rbrace$, this condition leads to {(cf.~\cref{eqn:knudsen-number})}
\begin{equation}
   \sbm =  \frac{\sqrt{c}}{\left({6 \mathrm{G} \mbh \rsp^3}\right)^{1/4} \rhosp } .
   \label{eqn:max-sig}
\end{equation}
The maximum accretion rate $\dot{M}_{\rm max}$ in terms of the Eddington accretion rate then can be approximated from \cref{eqn:accretion-rate-3b4} to be\footnote{For simplicity, we extrapolate the simple linear behaviour valid within the LMFP regime, which slightly overestimates the actual accretion rate, but is sufficient for the precision we aim for.}, 
\begin{equation}
 \frac{\dot{M}_{\rm max}}{\medd} \sim \SI{3e-15} \beta \left({\frac{\mbh}{\unit{\solmass}}}\right)^{1/4} \left(\frac{\rhosp}{\unit{\solmass\per\kpc\cubed}}\right) \left({\frac{\rsp}{\unit{\kpc}}}\right)^{3/4}.
 \label{eqn:mdot_max_medd}
\end{equation}
{This suggests that for high-density spikes, such as in the system J1205$-$0000 \citep{fengDynamicalInstabilityCollapsed2022}, where at the time of the formation of the seed the central density is $\rho_0 \sim \SI{1e19}{\solmass\per\kpc\cubed}$, the accretion can be super-Eddington. However, in that particular scenario the high-density spikes will be in the SMFP regime, such that \cref{eqn:mdot_max_medd} does not directly apply. Therefore a quantitative analysis in this case would require a self-consistent formation simulation of DM spikes, which we defer to future work.}
\section{Discussion and Conclusion}
\label{sec:conclusion}

In this paper, we investigated the evolution of the DM profile as well as the accretion rates of DM spikes around SMBHs. We found that the spike profile quickly reaches a quasi-steady state consistent with expectations from analytical arguments. We also found that for small self-interaction cross-sections (corresponding to the LMFP regime), the accretion rate increases linearly with the same, reaches a maximum in the IMFP regime and then drops again towards very large cross-sections.

The maximum accretion rate possible in a DM spike depends on spike profile parameters, which in turn depend on $\sbm$. In addition, if we assume that a CDM spike precedes an SIDM spike during the formation phase, then during the phase of transitioning from a CDM to SIDM spike, the accretion rates can be Super-Eddington for large spike densities.

There are a few caveats that come with our modelling approach: Firstly and most obviously, we assumed Newtonian dynamics, even in regions close to the SMBH. It would not be too hard to include general relativistic corrections  in a post-Newtonian treatment. In the appendix, we have estimated the magnitudes of the first post-Newtonian terms. However, in this work, we wanted to understand the behaviour of accretion rates as a function of $\sbm$, given the Newtonian SIDM spike model \citep{shapiroSelfinteractingDarkMatter2014}. 
To also take into account relativistic effects, a next step would be to implement DM self-interactions in an SPH-based numerical relativity code \citep{rosswogSPHINCS_BSSNGeneralRelativistic2021}.

In this work we simulate SIDM with a velocity-independent isotropic cross-section. In contrast, often a velocity-dependent cross-section is assumed to address the $\Lambda$CDM small scale structure problems while at the same time avoiding strong constraints at high velocities. However, in the context of the DM spike we find that also significantly smaller and therefore unconstrained cross-sections can be relevant. In fact velocity-independent scatterings are generally even much less constrained in the current context as the velocity dispersion in the DM spike is typically very large compared to other astrophysical systems. Also, note that simple light mediator models which would result in a velocity dependent cross-section are severely constrained by complementary probes~\citep{Bringmann:2016din,Kahlhoefer:2017umn}. On the other hand it is very easy to realise a well-motivated velocity independent scenario from a particle physics perspective, see e.g.~\cite{Bringmann:2022aim}. 

In summary, we have performed the first $N$-body simulations of SIDM spikes around SMBHs and studied the dependence of the accretion rates on the velocity-independent self-interaction cross-section.
\begin{itemize}
    \setlength\itemsep{0.5em}
    \item The density profile of the DM spike always converges to a profile with a spike index $\gamma\approx3/4$ regardless of the spike index of the initial power-law profile.
    \item We find that the accretion rate grows linearly with the self-interaction cross-section in the LMFP regime, which can be explained by a simple analytical model. A calibration of the free parameter, $\beta$, in the gravothermal fluid model for DM spikes, can be found in appendix~\ref{app:accretion}. In addition, we also observe that the accretion rates scale non-linearly with the cross-section in the IMFP regime. In both  regimes, our simulations match analytic expectations given the nature of conductivity of SIDM. The accretion rates can also be super-Eddington for reasonable spike parameters.
    \item These increased accretion rates could be incorporated in the subgrid models of BHs in \nbody codes with which cosmological simulations of SIDM are performed.
\end{itemize}

\begin{acknowledgements}
SVM and MB thank Stephan Rosswog for helpful discussions on numerical relativity. SVM thanks Sowmiya Balan and Felix Kahlhoefer for discussions.
MSF is delighted to thank Rainer Spurzem for a helpful discussion on the gravothermal fluid model. All authors thank members of Darkium SIDM Journal Club for discussions.
We acknowledge funding by the Deutsche Forschungsgemeinschaft (DFG, German Research Foundation) under Germany's Excellence Strategy -- EXC 2121 ``Quantum Universe'' -- 390833306.
MSF acknowledges support by the COMPLEX project from the European Research Council (ERC) under the European Union’s Horizon 2020 research and innovation program grant agreement ERC-2019-AdG 882679.
The simulations have been carried out on the computing facility Hummel (HPC-Cluster 2015) of the University of Hamburg.
Preprint numbers: DESY-25-076. 
\end{acknowledgements}

%
%

\bibliographystyle{aa} 
\bibliography{main} 

\appendix
\section{Post-Newtonian corrections}
\begin{figure}
    \centering
    \includegraphics[width=0.45\textwidth]{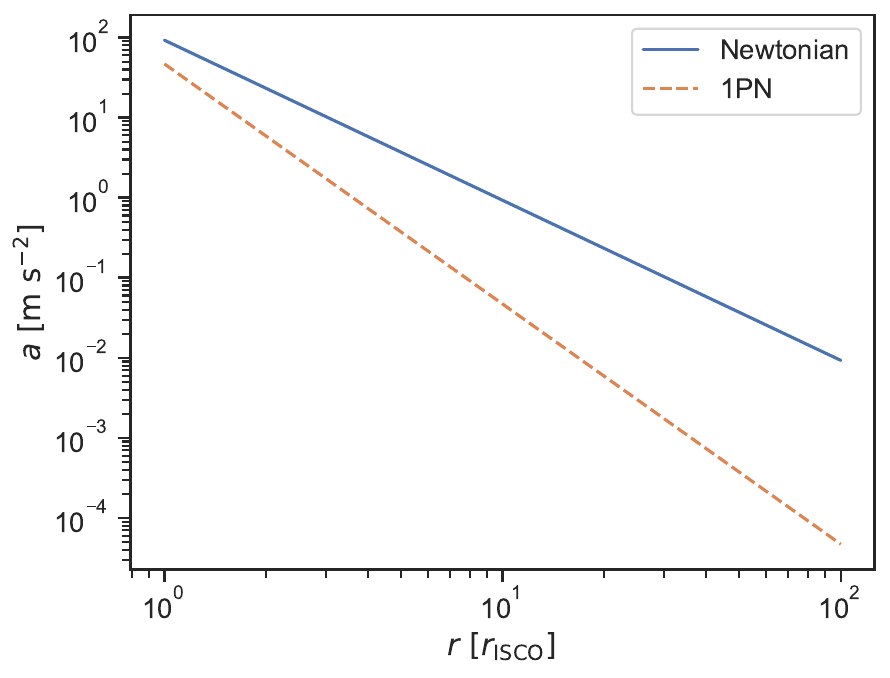}
    \caption{Newtonian acceleration, and 1PN contribution to the acceleration, plotted against distance of the test particle from the SMBH. The solid line and dashed line corresponds to the Newtonian and 1PN term, respectively.}
    \label{fig:newton-vs-pn}
\end{figure}

Post-Newtonian corrections to the orbital evolution are neglected in our simulations. 
We compare the magnitude of the correction to the Newtonian acceleration \autoref{fig:newton-vs-pn}. The relative acceleration of a binary at 1PN is given by \cite{poissonGravity2014},
\begin{align}
    \mathbf{a} = -\dfrac{\mathrm{G} M}{r^2} \mathbf{\hat{r}} &- \dfrac{\mathrm{G} M}{c^2 r^2}\left[ 
    (1+3\eta)v^2 - 3/2\eta\dot{r}^2 -2(2+\eta) \dfrac{\mathrm{G} M}{r} 
    \right] \mathbf{\hat{r}} \nonumber \\ &-\dfrac{\mathrm{G} M}{c^2 r^2} \left[2(\eta+2)\dot{r}v\right]\mathbf{\hat{v}}.
\end{align}
Here, $M=\mbh + m$, $\eta = \mbh m / M^2$ with $m$ being the mass of the orbiting particle.  
The above expression is weakly dependent on $m$ for any $m\lll\mbh$, in other words $M\approx\mbh$ and $\eta\approx0$.
The variables $r$, and $\dot{r}$ can be expressed in terms of the orbital elements namely, semi-major axis $a$, eccentricity $e$ and true anomaly $f$ as, $r=a(1-e^2)/(1+e\cos(f))$ and $\dot{r} = \sqrt{\mathrm{G} M/(a(1-e^2))} e \sin(f) $. 
We compare the magnitude of the Newtonian acceleration to the magnitude of the 1PN contribution in 
\cref{fig:newton-vs-pn}. 
Since we are only interested in order of magnitude estimate, we choose $e=0$ for convenience. 
Therefore, the tangential 1PN correction is initially zero.
The $x$-axis denotes the initial separation distance in units of $\risco$ of the primary and
the $y$-axis the acceleration. 
For $r \sim \risco$, the 1PN correction is subleading but nevertheless numerically relevant compared to the Newtonian force. Nevertheless, to obtain a first estimate and also facilitate a comparison with results in the literature,
we neglect these relativistic corrections and only study the Newtonian SIDM spike model in this paper.

\section{Eddington-Inversion}

Given a density profile, $\rho(r)$, the corresponding velocity distribution can be obtained by inverting the following equation,
\begin{equation}   \rho(r)=4\uppi\sqrt{2}\int_{0}^{\Psi(r)}f({\mathcal{E}})\sqrt{\Psi(r)-{\mathcal{E}}}\,\mathrm{d}{\mathcal{E}}.
   \label{eqn:density-profile-from-f}
\end{equation}
Here, $\Psi$ is the negative of the gravitational potential, $\mathcal{E} = \Psi(r) - v^2/2$ is the negative of the relative specific energy. \Cref{eqn:density-profile-from-f} is  recast from of $\rho(r) \propto \int f(r,v) \dd^3 \mathbf{v}$. Inverting \cref{eqn:density-profile-from-f}, we obtain
\begin{equation}
    f({\mathcal{E}}) = \dfrac{1}{\sqrt{8}\uppi^2} \dfrac{\dd}{\dd {\mathcal{E}}} \int_0^{\mathcal{E}} \dfrac{\dd \Psi}{{\mathcal{E}}-\Psi} \deriv{\rho}{\Psi} .
    \label{eqn:eddington-inversion}
\end{equation}
In our case, $\Psi(r) = \mathrm{G} M_{\rm BH} / r$. Therefore, the density profile becomes,
\begin{equation}
    \rho(\Psi) = \rhosp \left(\dfrac{\Psi/}{ \mathrm{G} M_{\rm BH}\rsp}\right)^{-\gamma}  \left(1+\left(\dfrac{\Psi/}{\mathrm{G}M_{\rm BH} r_t }\right)^{\beta}\right)^{-\alpha} .
    \label{eqn:truncated-spike-profile-potential}
\end{equation}

Using \cref{eqn:truncated-spike-profile-potential} in \cref{eqn:eddington-inversion}, and after some algebra, we compute the integral analytically using Mathematica for a fixed $\beta$. Given $f(\mathcal{E})$, we compute the $f(v)$ as follows,
\begin{equation}
    f(v,r) = 4 \uppi v^2 \dfrac{f(\mathcal{E}(v,r))}{\rho(r)},
\end{equation}
here $f(v)$ is evaluated for every sampled particle whose position is $r$. Suppose we have a particle $i$ at $r_i$, whose magnitude of velocity is $v_i$. To compute $v_i$, we first find the cumulative distribution function $F(v)$ from the distribution function $f(v,r_i)$. Then we employ the inverse transform sampling method to find $v_i$. That is, we draw a uniform random variable $u$ in the interval $[0,1]$, and then we set $v_i = F^{-1}(u)$.

\section{Accretion rate}
\label{app:accretion}
In this appendix we provide additional details on the analytical treatment of the accretion rate.
We start by reviewing the relation between the luminosity $L$ and the accretion $\dot{M}$.
The total energy of a particle in the gravitational potential of a SMBH is given by $E=-\mathrm{G}\mbh m /r$. Therefore, the differential energy change across a small distance $\dd r$ is $\dd E = \dd r \mathrm{G} \mbh m / 2r^2$. 
In SIDM spike, the DM particles transport energy from the centre. Therefore, as the mass $m$ loses an energy of $\dd E$, it moves in towards the centre by $\dd r$. If $m=\dot{M} \dd t$ is the small amount of mass that falls in, then $\dd E = \dd r \mathrm{G} \mbh \dot{M} \dd t / 2r^2$. Here, the luminosity represents this radial transfer of energy, meaning $\dd E/ \dd t = \dd r \mathrm{G} \mbh \dot{M} / 2r^2$. Integrating from $r=\risco$ to $\infty$, we get
\begin{equation}
 \   L = \dfrac{\mathrm{G} M_{\rm BH} \dot{M}_{\rm BH}}{2 \risco} .
    \label{eqn:newtonian-accretion}
\end{equation}
This relation has also been established in \cite{pringleAccretionDiscsAstrophysics1981,frankAccretionPowerAstrophysics2002,carrollIntroductionModernAstrophysics2017}.

{In the following we will estimate the accretion rate within the LMFP regime analytically.} 
From \cref{eqn:newtonian-accretion}, we have
\begin{align}
    \dot{M} = \dfrac{2 L_{\rm emp}\risco}{\mathrm{G} \mbh} ,
    \label{eqn:mdot-emperical}
\end{align}
where we have used $L_{\rm emp}$ to indicate that we are modelling luminosity using empirical relations. The empirical relation for the conductivity $\kappa_{\rm lmfp}$ in the LMFP regime is \citep{yangGravothermalSolutionsSIDM2023,kodaGravothermalCollapseIsolated2011a},
\begin{equation}
    \kappa_{\rm lmfp } = \dfrac{3}{2}\beta \rho(r) \dfrac{H^2(r)}{t_{\rm SIDM}} ,
\end{equation}
where $\beta$ is a free parameter that is calibrated against \nbody simulations. It is expected that $\beta$ is of order unity in the range \numrange{0.5}{1.5} \citep{essigConstrainingDissipativeDark2019,yangGravothermalSolutionsSIDM2023}. The above expression is the same as \cref{eq:conductivity-lmfp} but with the appropriate empirical coefficient. Substituting the above expression in \cref{eqn:heat_flux}, we get
\begin{align}
    L_{\rm emp} =  -m_\chi \dfrac{3}{2}\beta \rho(r) \dfrac{H^2(r)}{t_{\rm SIDM}} \dfrac{\partial \sigma_{1D}^2}{\partial r} .
\end{align}
In the innermost regions for $r\ll r_t$, the truncated density profile given in \cref{eqn:truncated-spike-profile} can effectively be replaced by the spike profile given in \cref{eqn:spike profile}. Substituting \cref{eqn:spike profile,eqn:relaxation timescale,eqn:velocity-dispersion-spike} and $H\approx r$ \citep{shapiroSelfinteractingDarkMatter2014} in the above equation, we get,
\begin{equation}
    L_{\rm emp} = 6 \uppi  \beta  \rhosp^2 \sbm \left(\frac{r}{\rsp}\right)^{-2 \gamma } \left(\frac{\mathrm{G} \mbh r}{\gamma +1}\right)^{3/2} .
    \label{eqn:luminosity-spike-emp}
\end{equation}
    For $\gamma=3/4$, we see that $L_{\rm emp}$ no longer depends on $r$. Correspondingly, the accretion rate from \cref{eqn:mdot-emperical} becomes,
    \begin{equation}
        \dot{M} = \frac{576 \uppi  \beta \rhosp^2 \sbm (\mathrm{G} \mbh \rsp)^{3/2}}{7 \sqrt{7} c^2},
        \label{eqn:accretion-rate-3b4}
     \end{equation}
 where $c$ is the speed of light. Substituting the parameters from \cref{tab:cdm-7b3-oj287}, we get
\begin{equation}
    \dot{M} = \beta \ \SI{3.8}{\solmass\per\yr} \dfrac{\sbm}{\unit{\cm\squared\per\g}} .
\end{equation}
Thus, we recover our fitted value $B^{-1}\approx\SI{3.7}{\solmass\per\yr\per\cm\squared\g}$ using $\beta=0.97$.

\section{Energy conservation}\label{sec:energy-conservation}
\begin{figure}[H]
    \centering
    \includegraphics[width=0.45\textwidth]{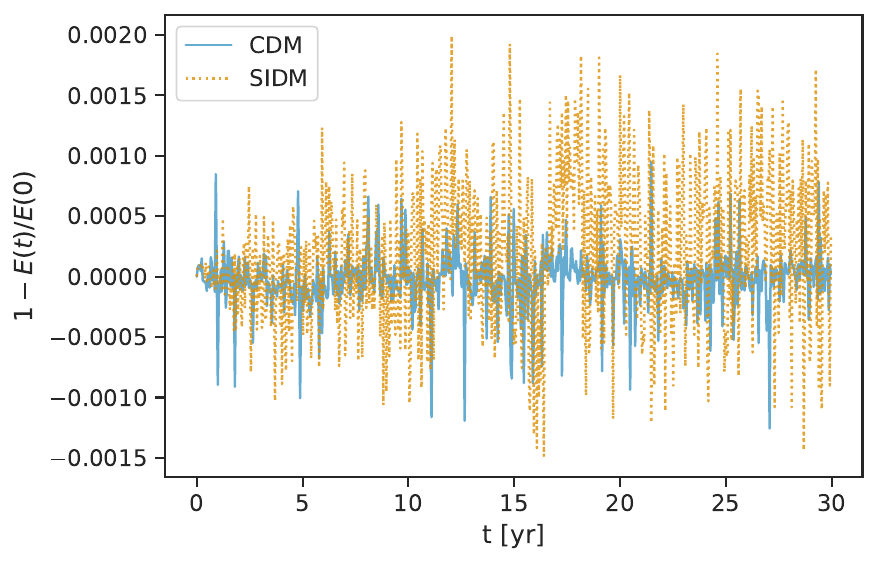}
    \caption{Fractional energy change over time is plotted. Solid line corresponds to CDM, and dotted line corresponds to a SIDM simulation with $\sbm=\SI{2.0}{\cm\squared\per\g}. $}
    \label{fig:energy-conservation}
\end{figure}
We verify our choice for the precision parameters by measuring how well the total energy is conserved in the simulations. 
For this test, we run a CDM and an SIDM simulation with $\sbm=\SI{2.0}{\cm\squared\per\g}$, and we do not remove particles that cross the $\risco$ of the SMBH. The parameters of the profile used in the simulation are given in \cref{tab:cdm-7b3-oj287}, and the precision parameters in \cref{tab:precision}. We present the results in \cref{fig:energy-conservation} and we see that total energy is conserved to within 0.1\% and 0.2\% in the CDM and SIDM simulations respectively.

\end{document}